# LOCAL CHANGES IN THE TOTAL ELECTRON CONTENT IMMEDIATELY BEFORE THE 2009 ABRUZZO EARTHQUAKE

by

P. I. Nenovski[1], M. Pezzopane[2], L. Ciraolo[3], M. Vellante[4], U. Villante[4], M. DeLauretis[4]

[1] University Center for Space Research and Technologies, Sofia University, Sofia, Bulgaria
[2] Istituto Nazionale di Geofisica e Vulcanologia, Rome, Italy ,
[3] ICTP, Trieste, Italy,
[4] Dipartimento di Fisica e Chemique dell'Università dell'Aquila, L'Aquila, Italy,

**Abstract.** Ionospheric TEC (Total Electron Content) variations derived from GPS measurements at 17 stations before and after the 2009 Abruzzo earthquake (EQ) of magnitude Mw6.3 were processed and analyzed. The analysis included interpolated and non-interpolated TEC data. Variations in the TEC of both regional and local characteristics were revealed. Several regional changes were observed in the studied period: 1 Jan-21 Apr. 2009. After analyzing non-interpolated TEC data of 5 GPS stations in Central Italy (Unpg (Perugia), Untr, Aqui (Aquila), M0se (Roma) and Paca (Palma)), a local disturbance of TEC was also found. This local TEC disturbance arises preparatory to the EQ main shock occurred at 01:32 UT on 06 April 2009, maximizes its amplitude of ~ 0.8 TECu after the shock moment and disappears after it. The TEC disturbance was localized at heights below 160 km, i.e. in the lower ionosphere.

**Corresponding author:**

Dr Petko Nenovski

## 1. Introduction

Electromagnetic perturbations due to seismic activity have been known for a long time (Milne, 1890). Variations in ionospheric parameters above seismically active regions are one of most actual aspects of these perturbations. Since the pioneering investigations devoted to ionospheric effects of the powerful Alaska earthquake occurred on March 28, 1964 ($M = 9.2$), extensive research of seismic-related anomalous effects in different ionospheric parameters has been carried out for a few decades (Davies and Baker 1965, Barnes and Leonard 1965, Datchenko *et al*. 1972, Larkina *et al*. 1983, Gokhberg *et al*. 1983, Parrot and Mogilevsky 1989, Liperovsky *et al*. 1992, Hayakawa 1999, Hayakawa and Molchanov 2002, Strakhov and Liperovsky 1999, Pulinets and Boyarchuk 2004). Among all the ionospheric parameters being sensitive to strong earthquakes are variations in the F2 region and the Total Electron Content (TEC). Variations in the F2 region parameters have been frequently revealed a few days before strong earthquakes by means of ground-based vertical sounding (Gokhberg *et al*. 1988, Gaivoronskaya and Zelenova 1991, Pulinets 1998, Ondoh 1998, 2000, Liu *et al.* 2000, Silina *et al*. 2001, Rios *et al*. 2004). A decrease of $fo$F2 from its monthly median at single ionosonde station Wakkanai is observed within ±3 days around the strong earthquake with M = 7.8 in Japan (Ondoh, 1998, 2000). Decreases of $fo$F2 observed at 1st, 3rd and 4th days before the main shock of the powerful Chi–Chi earthquake at single ionospheric station in Taiwan (M = 8.2) also have been recorded by Liu et al. (2000). They have found that the corresponding electron density decrease is about 51% from its normal value obtained from 15-day median process. The $fo$F2 decreases are seen between 12:00 and 18:00 LT. Very close similarities in most parameters describing the precursory anomalies (leading time, sign of $fo$F2 and value of electron density depletion, duration of each anomaly, and time period seen in LT) have been considered by Hobara and Parrot (2005). Simultaneous records from 60 different ionospheric stations have enabled Hobara and Parrot to separate the global events and local ones, and to find that for Hachinohe EQ event (M = 8.3) a *fo*F2 decrease down to 3MHz 4 days prior to and in 2 days after the earthquake have been observed. This effect takes place in the course of one day only and maximizes in the afternoon (15:00 LT) hours. Later, statistical analyses have been conducted on possible relationships between the *fo*F2 effects and 184 earthquakes with M > 5.0 occurring during years 1994÷1999 in the Taiwan area (Liu et al., 2006). Liu et al (2006) have revealed that the effect of *fo*F2 decrease (by > 25%) takes place afternoon time and within 5 days before the earthquake. Moreover, this effect increases with the earthquake magnitude but decreases with the distance from the epicenter to the ionospheric station. Liu et al (2006) have yet pointed out that only the M > 5.4 earthquakes have a significant chance to result in the mentioned *fo*F2 decrease and only those of them which were within the distance of 150 km.

Statistical analyses on ionospheric changes prior to strong earthquakes show that obvious abnormal TEC disturbances occur around the epicentral area (of hundreds and even thousand km) several days before the occurrence of earthquakes (EQ) (Liu et al, 2,000, 2001, 2004). It is not surprising that the vertical *TEC* (total electron content) obtained with using GPS (dual frequency measurements) is also very sensitive to changes in the *foF2* electron density measured by ionosondes. According to Houminer and Soicher (1996) the correlation between *TEC* and *foF2* can reach the value of 0.9. In that way, the anomalous ionosphere modification before some strong earthquakes of different Earth regions has been found out with using GPS TEC measurements in the recent years (Calais and Minster 1995, Liu *et al*. 2002, 2004, Plotkin 2003, Pulinets *et al*. 2005, Krankowski *et al*. 2006, Zakharenkova *et al*. 2006, 2007a,b, Ouzounov et al (2011), etc.). Results from TEC measurement around Chi–Chi earthquake by Liu et al. (2001) demonstrate severe depletion region of TEC around the epicenter (with a radius of 100–200 km) at some days before the earthquake. A 15-day running median of the TEC and the associated inter-quartile range have been utilized as a reference for identifying abnormal TEC signals during 20 M ≥ 6.0 earthquakes in the Taiwan area from September 1999 to December 2002 (Liu et al, 2000, 2004). Their results show that the pre-earthquake TEC anomalies appear during 18:00–22:00LT within 5 days prior to 16 of all the 20 M ≥ 6.0 earthquakes. Whereas the satellite measurements by Pulinets et al. (2001) for several seismic events located at various latitudes show either a localized enhancement or a decrease of electron density with spatial extent about 20° in latitude and longitude. One day before the Kythira (Southern Greece) earthquake occurred on 8 January 2006, a significant increase of *TEC* at the nearest stations, up to the value of 50% relative to the background condition and existing from 10:00 till 22:00 UT has been recorded; The area of this significant *TEC* enhancement has a size of about 4000 km in longitude and 1500 km in latitude (Zakharenkova et al, 2007a). Seismo-ionospheric anomalies in GPS TEC over European and Japan regions have been analyzed by Zakharenkova *et al.* (2007b) and provoked them to conclude that the occurrence of such variations may be registered in Europe 1-2 days prior earthquakes, while for very strong Japanese earthquakes this temporal interval can reach 5 days. The GPS/TEC data indicate an increase and variation in electron density reaching a maximum value on March 8, 2011 – 3 days before the Mw9 Tohoku EQ (Ouzounov et al, 2011).

Recently an enhancement of ionospheric Total Electron Content (TEC) *immediately* before the 2011 Tohoku-oki earthquake ($M_w$9.0) has been reported by Heki (2011). The TEC enhancement emerges ~ 40 minutes before the main shock. Later, Heki and Enomoto (2013) have scrutinized the nature of characteristics of the TEC change preceding the 2011 Mw9 Tohoku earthquake. The authors first have confirmed the reality of the enhancement using data of two other sensors, ionosonde and magnetometers. The amplitude of the preseismic TEC enhancement is within the natural variability, and its snapshot resembles to large-scale traveling ionospheric disturbances. However, distinction could be made by examining their propagation properties. Second, similar TEC anomalies occur before all the $M \geq 8.5$ earthquakes in this century, suggesting their seismic origin (Heki and Enomoto, 2013).

In this paper using data from 17 GPS stations spread over Italy and Greece we thoroughly analyze both temporal and spatial characteristics of ionospheric TEC variations in association with the 2009 Abruzzo earthquake. We pay attention on TEC changes around the main EQ shock occurred on 6 April 2009. We differentiate regional changes from local ones and then juxtapose the observed local TEC changes with the recent Heki's findings.

**2. Data and analysis**

The very destructive, Abruzzo earthquake occurred close to L'Aquila on April 06, 2009, 01:32 UT. The geographic coordinates of the epicenter was 42.33N and longitude 13.33E and its magnitude – Mw = 6.3 ($M_L$ = 5.8). According to the Istituto Nazionale di Geofisica e Vulcanologia (INGV), this earthquake was classified as an $M_L$ = 5.8 event, with a depth of 8.8 km. It was preceded by a persistent seismic activity for approximately three months: namely, between January 16 and April 5, 2009, 34 seismic events with $2 < M_L < 3$ (and 9 with $M_L > 3$) was registered in the territory. The strongest event was followed by a large numbers of aftershocks with remarkable events on April 7, 17:47 UT ($M_L$ = 5.3) and on April 9, 00:52 UT ($M_L$ = 5.1) (Figure 1).

GPS system consists of more than 24 satellites, distributed in 6 orbits around the Earth at an altitude of ~ 20 000 km. Each satellite transmits dual very high frequencies of signals, 1575.42 and 1227.60 MHz. Ionospheric TEC can be computed on the basis of phase delay between Global Positioning System (GPS) station's dual frequencies while electromagnetic wave propagates through ionosphere. The oblique (slant) TEC, the integral of the electron density over a line of sight from a ground receiver to a satellite on the signal propagation path, can be estimated from the standard GPS observations (pseudo-range and phase, relative to the two available carriers $f_1$ (1575.42 MHz) and $f_2$ (1227.60 MHz). This is done forming the differential delays of the pseudo-ranges (directly) and phases (transformed into optical paths $L_1$ and $L_2$) relative to the two carriers. Properly combining the code and phase differential delays one gets the slant TEC (STEC) between a GPS satellite and a ground based dual-frequency receiver, which can be written as

$$\text{STEC} = a\,[f_1^2 f_1^2/(f_1^2 - f_1^2)][(L_1 - L_2) - (\beta_r + \beta_r) - \mu_{Arc}] \tag{1}$$

where $a = 1/40.3$, $\beta_r + \beta_r$ are the differential hardware biases for receiver and satellite, respectively and $\mu_{Arc}$ an additional term, variable from arc to arc, depending on the way the receiver processes pseudo-range. For the data used in present work, a calibration technique attempts to estimate, cumulatively, the hardware biases plus the term $\mu$ for each arc. Note presence of gaps in the data may severely affect the calibration. Unfortunately, important TEC data from the *Aqui* (L'Aquila) station (the closest to the earthquake epicenter) are interrupted around the EQ shock, and unfortunately need to be cancelled. Another caution comes from the fact that each satellite–receiver pair has a different measurement bias, therefore only their temporal changes are meaningful and analyzed further.

The slant TEC can be converted to vertical TEC (VTEC), which is the projection of oblique TEC on the thin-shell, using an elevation mapping function (Dautermann et al, 2007). The location of a recorded VTEC is defined as the intercept of the ray path of the GPS signal and the ionospheric height (accepted as a thin shell at height 400 km). This intercept is termed the ionospheric (pierce) point (PP). The VTEC's so far described can be interpolated in order to estimate the VTEC in locations different from the PP's, such as the TEC station.

Time resolution of the set of interpolated TEC data we use is 5 min, i.e. 288 values are generated per day for each TEC station. GPS data from January 1 to April 20, 2009 of 17 stations (in Italy mainly and Greece) are processed and then corresponding TEC time series obtained. These time series are given in TEC units (TECu), where 1 TECu = $10^{16}$

electrons/m$^2$. Because of satellite and receiver biases, $\beta_s$ and $\beta_r$, and $\mu$ as well as different pierce points, PP, the calculated interpolated TEC data from different satellites can differ and the TEC difference can reach 1-2 TECu. As for non-interpolated TEC data only temporal changes are meaningful and hence taken into account.

## *2.1. Interpolated TEC data*

In statistics, envelope method is mostly used to identify possible significance of disturbances. Under the assumption of normal distribution with mean $\mu$ and standard deviation $\sigma$ of TECs and if an inter-quartile range is assumed (e.g. Liu et al, 2004), the expected values of upper bound and lower bound of envelope are $\mu \pm 1.34\sigma$. If the observed TEC falls out of either the associated lower or upper bounds of such an envelope, it is declared at confidence level of about 82% that a lower or upper abnormal signal is detected. Li et al (2009) have used bounds $\mu \pm 2\sigma$, for that case the confidence level is of 95%. Thus, upper and lower bounds of TEC variations can be determined at different confidence levels. The mean for a sliding window, which is 4 days long, is assumed as background TEC. TEC variations and corresponding upper and lower bounds fixed at $\mu \pm 1.34\sigma$ were inspected for 5 TEC stations, *Unpg* (43.1N,12,4E), *Untr* (42,6N,12.7E), *Aqui* (42.4N,13.4E), *m0se* (41.9N,12.5E) and *Paca* (40.9N,14.6E). For convenience only the interval 31 March–7 Apr 2009 is illustrated (Figure 2). The four TEC stations: *Unpg*, *Untr*, *m0se* and *Paca*, are the closest ones to the EQ epicenter: with distances respectively ~110, ~60, ~90 and ~180 km from Aquila (Figure 1). The actual TEC variations (in TECu) of each TEC stations are in blue, while the upper and lower bounds are marked respectively with red and green lines. As it is seen there are two moments when the TEC value is below the $\mu - 1.34\sigma$ value (on 1-2 April) or exceeds $\mu + 1.34\sigma$ (on 5-6 Apr). Inspecting the TEC variations for the whole interval, 04 Jan-21 Apr 2009, one can see (that) there are a lot of days (or intervals) on which the TEC values are definitely above the upper bound of $\mu +1.34\sigma$ for all 5 stations (such events are clearly indicated on 24 January; 01-02, 14, 21-22 and 27 February; 8-9, 21 and 24 March; and 09 April 2009 implying that positive anomaly variations occurred in these days (intervals). Variations at one TEC station only have also been observed: i) on 09 January at station *m0se* (close to Rome), ii) on 12 January at *Unpg*, and iii) at L'Aquila on April 5. TEC spikes on 09 and 12 January are false signals due to data gaps. An extreme anomalous TEC disturbance of 3 TECu is recorded on 5 April only at Aquila. Its duration is at least 11-12 hours.

Looking at these TEC interpolated data (Figure 2) two kinds of TEC disturbances were discriminated: i) TEC disturbances of *regional* character that appear simultaneously at all TEC stations in the L'Aquila area; and ii) disturbances of *local* character that emerge at only one TEC station. The causes of disturbances of the first class need to be sought in various regional and/or global factors, such as solar/geomagnetic activity, meteorological/lightning activity, etc. which can significantly contribute to ionosphere TEC variations. The ionosphere is definitely under the control of the solar and geomagnetic activity. For the 01 January-21 April 2009 the geomagnetic conditions were quiet and the geomagnetic index Kp practically was low (less than 2). Irrespectively of the quiet geomagnetic activity level, effects on TEC variations due to global and/or regional factors are still present. Note although the considered interval was geomagnetically quiet, still there were some geomagnetic storm sudden commencement (SSC) events. The following SSC (minor) events are registered on: Jan 25 (22:24 UT), Feb 3 (20:12 UT), Feb 20 (20:12 UT), Mar 3 (06:02 UT), Apr 24 (00:53 UT) (for details see [ftp://ftp.ngdc.noaa.gov/STP SOLAR_DATA/SUDDEN_COMMENCEMENTS/STORM2.SSC.](ftp://ftp.ngdc.noaa.gov/STP SOLAR_DATA/SUDDEN_COMMENCEMENTS/STORM2.SSC.)

Figure 3 illustrates interpolated TEC variations at L'Auila area. The regional TEC data (from *Unpg* to *Paca*) before 3 April were fully coincident except an interval of increased dispersion on 3-6 April 2009. The average value of the cross-correlation coefficient of daily values of (*Unpg*, *Paca*) is high and equal to 0.984 ranging between 0.979 and 0.992. An operation failure of the GPS receiver (*Aqui* station) at L'Aquila however occurred at 02:25 UT on the EQ day (6 April 2009) − 53 minutes after the EQ shock moment. Because of TEC data interruption around the EQ shock moment and subsequent data calibration problems, changes of TEC recorded at L'Aquila station (42.4N, 13.4E) was clearly detached from the TEC trends observed at the other TEC stations in the Aquila area: *Unpg* (43.1N, 12,4E), *Untr* (42,6N, 12.7E), *m0se* (41.9N, 12.5E), and *Paca* (40.9N, 14.6E). TEC changes from *Aqui* around the EQ shock moment thus was considered as fictious and hence, will not be considered further.

In order to find local (meaningful) changes of TEC in time, regular changes in the TEC (diurnal ones) should be removed. The following quantity is introduced:

$$\text{DTEC} = (\text{TEC}(i,j) - \mu(\text{TEC})(k,j))>)/\sigma(\text{TEC})(k,j), \ i\text{-}5 \leq k \leq i\text{-}1 \qquad (1)$$

TEC value ($i$, $j$) as a function of day, $i$, and minute, $j$, ($j$ = 1 to 288) is exploited. $\mu$(TEC) and $\sigma$(TEC) denote mean value and the standard variation calculated over previous 4 days, $i$-5 ÷ $i$-1. Our chose corresponds to a 4-day running mean, which is enough to remove the variations larger than 4 days. It is worth noting that such a choice is dictated by the 5 minutes resolution of TEC data. The diurnal TEC variations are strongly dependent on and move forth/back with the sunrise/sunset time. The sunrise/sunset time rises (decreases) having a velocity ranging between zero and minute and so per day. Thus, the 5 minute resolution of TEC data confines the averaging procedure of mean TEC variations roughly to 4 days. The difference TEC($i$,$j$) − $\mu$(TEC)($k$,$j$) thus represents the TEC signal to be investigated for its possible relationship with earthquake activity. In (1) it is evaluated by comparison with the corresponding natural/observational TEC noises represented by $\sigma$(TEC)($k$,$j$), which describes the overall (local) variability of the signal including all sources of its variability, observed at daytime moment $j$ in similar observational conditions occurred on previous $k$ days. In this way, the measured TEC signal can be quantified in terms of signal to noise (S/N) ratio. DTEC (1) is henceforth called *TEC index* (or TEC). Calculation of DTEC variations according to (1) means that we consider the TEC variations as signals of Gaussian distribution. For standard Gaussian distributions of signal the mean of DTEC should be zero. If an anomalous signal however exists in the time series, it is expected to emerge (clearly) detached from the Gaussian distribution. Consequently the mean of such a signal should be away from zero. For theoretical foundation of signal detection problems we refer to the Neyman-Pearson test of statistical hypotheses (see Neyman and Pearson, 1933).

Further, daily TEC indices can be calculated. It is performed by averaging over the all 288 TEC index values per day. Applying daily TEC index (mean of (1)) we are thus able to discriminate possible anomalous signal from TEC data series. The determined *daily* TEC index variations for the four stations in the L'Aquila (*Unpg*, *Untr*, *m0se*, *Paca*) area are disposed for the period 01 Jan–21 April 2009 (Figure 4). One sees that the daily TEC indices behave similarly and executes several coinciding extreme for all 4 TEC stations indicating TEC anomalies *of regional type*. In the studied period such events appear at least on 1-2 February, 8-9 March and 5-6 April 2009. On 5 April 2009 a distinct (regional) increase of TEC density however covers all latitudes from *Tori* (Torino) to *Paca* at least (over 5 degrees in latitude (see Figure 5).

We remind that TEC increases at stations *Unpg, Untr, m0se* and *Paca* (the L'Aquila area) appeared on day before the EQ day (6 April) could be considered as TEC background (of non-local character) for a further analysis of non-interpolated TEC data on EQ days (5-6 April 2009).

## 2.2. Non-interpolated data

Interpolated TEC data do not help to identify TEC disturbance of local extent. The main reason is that the PPs from one GPS station are widely distributed and are intermixed with the PPs of other GPS stations. Therefore actual (non-interpolated) GPS TEC data with sampling frequency of 30 seconds from each GPS satellite are also examined. Non-interpolated TEC data are obtained from all satellites that are over the horizon at given time *t* with respect to the questioned GPS receiver (the minimum elevation angle EL is roughly around 10° (degrees). In our analysis we exploit TEC data of satellites with EL greater that 67 degrees. Of course, during the course of day different satellites appear at a given GPS receiver and corresponding TEC data collected from each satellite with EL > 67° are of short duration (several tens of minutes and less). Figure 6 sketches sample elevation angle, azimuth angle and vertical TEC (VTEC) changes over ~6 hours period observed at Aqui station with the satellite #8. In addition, VTEC data from satellite #8 at *Untr* are put on that at *Aqui*. The TEC shows gentle curvatures due to satellite elevation changes. The VTEC trends at *Aqui* and *Untr* (~50-60 km distance between them) are practically coincident.

Figure 7 represents pierce points (PPs) of GPS satellites referred to *Untr* (in black) and *Aqui* (in blue) stations. The EQ epicenter is given by red star. The pierce points on 6 April are calculated for elevation angles exceeding 70 degrees, so pieces of GPS satellite trajectories projected/mapped as pierce points at 300 km height, are sketched. A crossing of pierce points of the two stations is observed The TEC variations at the two stations might be identical provided that the spatial scales of TEC exceed considerably the distance between *Untr* and *Aqui* (an assumption). In order to avoid possible intersection of pierce points (PP) of the two GPS stations (Figure 7), say *Aqui* and *Untr*, elevation angle (EL) should be increased, e.g. extreme EL > 86° would provide such a separation. Then Pierce Points (PPs) of given GPS station would lie within a circle of radius less than 30 km centered above the GPS receiver. Choosing much higher elevation angles, e.g. extreme EL > 86° suggests that non-interpolated TEC data will not provide continuous set of data points. This circumstance will produce numerous data gaps (typically for TEC data gathered from GPS satellites). Namely, such TEC data gaps do not allow to record uninterruptedly the whole evolution of the disturbance process occurred around the earthquake moment above the EQ epicenter and in principle, detection of the utmost (peak) amplitudes of the TEC disturbance processes may be realized accidentally. As can be seen, the non-interpolated TEC data are grouped and each group (spot centered, or located in time) contains data (30 points in average) only from one satellite being over the questioned GPS stations.

Figures 8a and 8b were drawn where 24 hour variations in TEC (non-interpolated data, all satellite data ) at *Untr* and *Aqui* are plotted. The TEC data are practically coincident except satellites #8, #9 and #29. Three satellites #8, #9, and #29 (marked with ellipses) are shown because of indicating TEC differences of up to several TECu. An increase of TEC on 5-6 Apr (of ~3 TECu) with respect to 4-5 April (of ~ 2 TECu) at *Untr* was detected at midnight hours. These differences however are not considered as reliable because of the data interruption at Aqui station on 6 April. Therefore, the TEC increase at *Aqui* (as by recorded satellites #8, #9, #29) thus is considered doubtful. Further, *Untr* TEC data were used as indicative of possible local TEC variations expected over *Untr* and *Aqui* stations separated at a distance ~50-60 km. On the other hand, the two *Untr* and *Aqui* GPS stations are at distances of ~ 90 km from the *m0se* GPS station. TEC differences (of local character) occurred between two GPS stations in the L'Aquila area – *M0se* and *Untr* thus are further analyzed.

A TEC difference method is suggested here based on consecutive satellite TEC data at two TEC stations. Differences $\Delta TEC = TEC_{Aqui} - TEC_{untr}$ and $\Delta TEC = TEC_{untr} - TEC_{m0se}$ of non-interpolated TEC data on days 28 March - 06 April 2009 are constructed for stations *Aqui* and *Untr* (not presented here) and for stations *Untr* and *m0se* (*Unpg*). TEC data at given time *t* from each satellite with elevation angle EL exceeding simultaneously a certain value as regards to two stations are successively substracted to each other. This method will allow TEC values from pierce points sets to be detached to each other.

One sees that with one exception (to be examined later) the ΔTEC variations between different satellites rarely exceed 0.2 TECu (outliers). The absolute error of TEC measurements is 0.01 TECu ($10^{14}$ electrons/m$^2$). The standard deviation varies from case to case and in fact lies between 0.05 and 0.145. Unfortunately, TEC data from satellites with EL exceeding 86° are lacking around the EQ shock moment. GPS satellites with EL > 86° were absent between 16:00 UT (on 5 April) and 04:50 UT (on 6 April) for the Aquila area. In seeking of non-interpolated TEC data that would cover the EQ shock moment the elevation angles EL was reduced to 67° Interestingly, TEC differences from satellites with EL > 67° revealed definitely different behavior − a hump-like distribution of the ΔTEC difference appears centered close to the EQ shock moment.

Its amplitude exceeds well the noise level and standard deviations that were already determined; the obvious hump-like distribution centered close to the EQ shock moment also works against possible accidental character of this positive ΔTEC difference. Hence, the positive ΔTEC difference suggests that PP area at *Untr* and *m0se* would be thus definitely detached to each other (Figure 9b). For EL > 67° this would occur at heights less that 160 km. Hence, a possible explanation is that the observed ΔTEC difference would be produced at lower heights, i.e. somewhere in the E layer.

The amplitude of the positive ΔTEC difference is the difference between two stations − *Untr* and *m0se*. This hump-like distribution is of amplitude 0 .3 ÷0.4 TECu prior to the EQ moment followed by a jump increase immeadiately after the Eq shock moment to ~ 0.8 TECu. Two different possible processes or mechanisms are assumed: i) a *relative increase* of TEC between *Untr* and *m0se* stations that starts at the beginning of day 95 (5 April) and persists with some steady magnitude, say $dA_1$ ($dA_1 = A_{\text{untr}} − A_{\text{m0se}}$), through the earthquake shock and some time after it. $dA_1$ can be considered as a one polarity (positive) TEC disturbance centered at Untr station; TEC difference between m0se and Unpg stations was tested and did not reveal similar anomaly.

The examined non-interpolated TEC data (with elevation angle EL > 67°) reveal an existence of positive TEC disturbance at E layer heights localized to the *Untr* area placed at distance ~ 90 km from *m0se*. Note that *Aqui* area is placed approximately at the same distance (90 km from *m0se*. Note the *Untr* −*Aqu*i line lies approximately parallel to the Appenine's fault system.

## 3. Discussion

Using GPS TEC data from 17 TEC stations we investigated ionospheric anomalies for the period 01 January-20 April 2009, including the Mw6.3 Abruzzo earthquake on 06 April 2009. For the mentioned period TEC changes of local character centered in time at the EQ shock moment was registered. The positive TEC difference was localized to the EQ area close to *Untr*. Another positive TEC increase of regional character was also recorded on 5 April with an amplitude peak close to the EQ epicenter (between Perigia (*Unpg*) and Palma (*Paca*)). This regional TEC increase starts ~ 16 hours before the EQ shock moment and covers a zone from Torino (at North) to Cagliari (at South). In our analysis the TEC data from the EQ TEC station: *Aqui* was left without attention. The reason was already pointed out: GPS data interruption causing calibration errors.

Among the last 4 GPS regional stations, TEC values at *m0se* and *Unpg* (*Paca*) stations were nearly identical, subtracting them, the obtained TEC differences do not exceed 0.2 TECu. (Figure 9a). With one exception: If TEC values at *Untr* are subtracted by TEC values at *m0se* (used as a reference), a distinct TEC anomaly of amplitude of 0.8 TECu is observed (Figures 9a and 9b). This transient positive anomaly of TEC is seen to start at 23 UT − ~2 and half hours before the earthquake shock.

The TEC anomaly is located at *Untr* station close to the epicenter. This anomaly disappears and TEC differences fall again within the usual error interval of 0.2 TECu after ~one hour after the EQ shock. This TEC increase at *Untr* is the only anomaly among the period we study. Further, the latitudinal/longitudinal position of the *Untr* station (42.6N, 12.7E) (where the positive TEC anomaly was observed) is in NW direction from the EQ epicenter and approximately overlaps with the local and faults (including the ruptured one) oriented in NW-SE direction.

The only previous finding of positive TEC anomaly which appears immediately before the EQ shock, is by Heki (2011), Heki and Enomoto (2013). The positive TEC anomaly appears ~ 40 minutes before the great (M9) Tohoku earthquake on 11 March 2011. This transient anomaly emerges and disappears simultaneously at several TEC stations placed at different distances from the EQ epicenter. Another wave-like TEC disturbances of smaller amplitude appears some time earlier and propagates with a speed close to acoustic one far away from the epicenter (Heki and Enomoto, 2013). Heki (2011) however has found similar positive TEC anomalies anticipating other strong EQs around the world.

As opposed to the Heki's finding of positive TEC anomaly (Heki, 2011; Keki and Enomoto, 2013), the positive TEC difference recorded immediately before the L'Aquila EQ possesses the following characteristics:

i) It is located at low heights (probably at E layer heights). This finding follows from a requirement of non-overlapping PPs of *Untr* and *m0se* stations. Non-overlapping PPs occurs for heights <160 km (Figure 10). Note the Heki's finding refers to positive TEC changes occurred at ionospheric F2 layer heights;

ii) It retains its positive value for ~ 3 hours. The relative amplitude (between *Untr* and. *m0se*) reaches a value of 0.8 TECu at its maximum. The local TEC disturbance around the L'Aquila EQ represented a hump-like distribution and returned to the background level (< 0.2 TECu) within a hour after its maximum. A sudden depletion effect as it happened after the Tohoku EQ (Heki, 2011) was not observed after the L'Aquila EQ shock moment;

iii) It appears in a localized area close to the EQ epicenter (*Untr* station is placed at ~60 km from L'Aquila); It is more correctly to say the spatial scales of this positive TEC disturbance is less than the distance between *m0se* and *Untr* stations; A density gradient mechanism might produce a density expansion from unknown source (perhaps fault zone?) with some velocity. This velocity should be low and thus TEC difference appears.

Other geophysical evidences of the Mw6.3 2009 Abruzzo earthquake have been reported so far. Thermal infra-red (TIR) emissions near tectonic lineaments of Central Italy have been identified in space-time correlation with Abruzzo earthquake epicenter between 30 March and 1 April. The authors' findings are that TIR anomalies are indicative for seismic events of medium and low magnitude as foreshock with $M_L = 4.1$ occurred on 30 March (Lisi et al, 2010). Radon emission starting to be intensified on 30 March as well as TEC (regional) increase (on 5 April 2009) has been already reported by Ouzounov et al (2009). The spatial and temporal characteristics of both TIR anomalies and radon emission seem not to be in compliance with local and temporal scales of the transient TEC disturbances recorded immediately (hours) before the Abruzzo earthquake.

Various physical mechanisms have been suggested so far to explain observed ionospheric variations associated with earthquakes. For example, quasi-electrostatic (QE) fields (Pierce, 1976) and electromagnetic fields (Molchanov et al., 1995) penetration mechanisms have been proposed. Gravity waves (GW) as an agent of ionospheric variations (mainly in the low ionosphere) are examined by Molchanov and Hayakawa (1998), as well. Ionospheric variations are also considered to be initiated by gas (radon) release from the crust above earthquake preparation region (Pulinets et al., 1994). Alpha decay of radon gas released from the crust can also ionize the atmosphere. They may change the electric resistivity of the lower atmosphere, which could disturb the global electric circuit and redistribute ionospheric electrons (Pulinets and Ouzounov, 2011). Due to the stress of the rocks, electric charges at the Earth's surface and electric currents in the atmosphere—ionosphere system could appear (Freund, 2003, 2004, 2008; Pulinets et al., 2003). It is worth noting that such electric charges and currents under stress in laboratory conditions already have been measured (Enomoto and Hashimoto (1990, 1992), Freund (2000, 2004, Takeuchi et al., 2006). Then electric field/current in the ionosphere and Joule heating could modify and/or redistribute the electron concentration/temperature in height. A model of ionospheric variations based on the effect of atmospheric electric current flowing into the ionosphere was proposed by Sorokin et al (2006). As a result plasma density in the lower ionosphere increases and formation of an anomalous, sporadic E layer is possible (Sorokin and Chmyrev, 2010). A sporadic E layer may be generated by discharge processes (Ongoh and Hayakawa, 2002), as well. It is worth noting that sporadic E layers and their dynamics successfully were studied recently by TEC measurements (Maeda and Heki, 2014).

A promising hypothesis to explain the observed anomalous disturbances in TEC (even if they occur at E heights) may thus be related to a seismogenic electric fields/currents action. More efforts however would be desirable both in modeling and in monitoring of preparatory and seismogenic processes in the Lithosphere-Atmosphere-Ionosphere (L-A-I) system and their effects not only in the ionospheric F2 region but also in the lower ionosphere in order to highlight and quantify the chain of processes resulting in anomalous TEC events.

## 4. Conclusion

In this paper we have examined temporal and spatial extents of TEC anomalous changes around the destructive Abruzzo earthquake occurred on 06 April 2009. The observed changes in TEC were of regional and local character. The former appeared repeatedly on the EQ day and before it, while the local one was observed on the EQ day. Temporal changes of the local TEC disturbance around the moment of a strong earthquake shock are reported here. A TEC difference method is suggested based on consecutive satellite TEC data at two TEC stations and requiring their pierce points sets to be detached to each other. The local TEC disturbance thus was found to lie at E layer heights (less than 160 km). A preparatory nature of local TEC changes accompanying EQ shock moment is thus evidenced for the Mw6.3 Abruzzo earthquake. The findings suggest: i) admissible connection of EQ shock and generation of local TEC disturbances at lower ionosphere heights, and ii) growth of positive TEC disturbances amplitude approaching the EQ shock moment attaining its maximum value close or after the EQ moment.

**FIGURE CAPTIONS**

**Fig. 1** Map of GPS stations (marked with black triangles) located in Italy. Encircled are stations *Unpg*, *Untr*, *Aqui*, *m0se* and *Pac*a. Thick points (in red) illustrate the epicenters of earthquakes of magnitude M >4 occurred in Central Italy for 01 January – 30 April 2009.

**Fig. 2.** TEC variations and upper and lower bounds for L'Aquila area. TEC variations and corresponding upper and lower bounds fixed at $\mu \pm 1.34\sigma$ are depicted for 5 TEC stations: *Unpg* (43.1N,12,4E), *Untr* (42,6N,12.7E), *Aqui* (42.4N,13.4E), *m0se* (41.9N,12.5E) and *Paca* (40.9N,14.6E). The actual TEC variations (in TECu) of each TEC stations are in blue, the upper and lower bounds are marked respectively with red and green lines.

**Fig. 3.** Vertical TEC trends and data dispersion. VTEC daily variations in Central Italy for 31 March–7 April 2009 are shown. Note a good coincidence of all VTEC trends for two intervals: before 3 April and after 6 April. VTEC data in Central Italy however indicate a scattering effect both in night and day hours for 3-6 April. The geomagnetic activity for the whole period was extremely low (Kp < 3). Hence, this unusual scattering is not associated with the geomagnetic activity.

**Fig. 4.** Daily TEC index variations. The daily TEC index (TEC) is calculated by averaging over the all 288 TEC index values per day. The daily TEC index variations for L'Aquila and the four stations in the L'Aquila area are disposed for the period 01 Jan–21 April 2009. As one sees the daily TEC index behaves similarly at all stations and executes several extreme coinciding for all stations. The daily TEC index regularly bounds between mean(TEC) ± 2*std(TEC). There are peaks of regional increase in TEC on 1-2 Feb, 8-9 Marc and 05 April 2009.

**Fig. 5.** Vertical TEC distribution in latitude. VTEC data from GPS stations spanning North (one station, Torino), Central (5 station) and South (Cagliari) Italy are attracted. The latitude position of each GPS stations is named and indicated by vertical lines (dot). VTEC distributions in latitude at every 8 hours on day, 5 April, are shown. VTEC trends in latitude at 22 UT on 4 and 9 April (see brown and blue thick lines) are also drawn being used as reference trends. At 22 UT on 5 April a regional increase of VTEC of amplitude ~2 TECu (see violet line) with respect to the 22 UT VTEC data on 4 and 9 April was clearly observed. At 10 and 16 UT on 5 April only s*light* increases of VTEC at Aqui station were registered.These VTEC increases were recorded only at Aqui station and might be questioned because of the GPS data interruption occurred around the EQ shock. Besides, the VTEC trend changes its slope – VTEC starts to increase from Torino to Perugia, i.e. it seems that a TEC maximum should exists and placed between Unpg (Perugia) and Palma (Paca). This event of VTEC increase was detected also at 00 and 02 UT on 6 April. Note the Aqui TEC data for the time interval: 22 UT, 5 Apr–02 UT, 6 Apr (considered as erroneous) are consciously/tentatively cancelled (see ellipse).

**Fig. 6.** Elevation angle, azimuth angle and VTEC data taken from satellite #8 at Aqui station are sketched. For comparison VTEC data from satellite #8 at Untr are put on that at Aqui. The VTEC trends at Aqui and Untr stations (~55 km distance between them) are practically coincident.

**Fig. 7**. Pierce points of GPS satellites referred to Untr (in black) and Aqui (in blue) stations. The EQ epicenter is given by red star. The pierce points on 6 April are calculated for elevation angles only exceeding 70 degrees, so pieces of GPS satellite trajectories projected/mapped as pierce points at 300 km height, are sketched. A crossing of pierce points of the two stations is observed. The TEC variations at the two stations might be considered identical provided that the spatial scales of TEC structures exceed considerably the distance between Untr and Aqui (an assumption).

**Fig. 8a.** All satellite TEC data on 5 and 6 Apr at Aqui and Untr stations. The two dash vertical lines (left panel) indicate data gap (no satellites with elevation angle exceeding 84 degrees at the GPS station). An increase of TEC at Untr (right panel)on 5-6 Apr (of ~3 TECu) with respect to 4-5 April (of ~ 2 TECu) was detected at midnight hours. Note the TEC increase at Aqui (satellites #8, #9, #29) are considered doubtful and are not used in our analysis. The TEC data are practically coincident except satellites #8, #9 and #29.

**Fig. 8b.** All satellite TEC data on 5 at Aqui and Untr stations: details. The twin dash vertical lines indicate data gap (no satellites with elevation angle exceeding 84 degrees at the GPS station). The TEC data are practically coincident except the time interval 18-24 UT. The corresponding TEC increases at Aqui (in red) are, as it was mentioned in Figure 7, considered doubtful. Non-interpolated TEC differences DTEC between *Aqui* and *Untr* stations.

**Fig. 9a.** Non-interpolated vertical TEC difference TEC(untr) −TEC(m0se) for 28 March–8 April taken from all satellites crossing GPS stations in Central Italy with elevation angle exceeding 67 degrees at the two stations. This difference is close to 0 (with a mean value of 0.024 TECu) and typical for TEC data inferred from the considered GPS stations in Central Italy. The only exception is a time interval (of several hours) around the EQ shock moment (marked with an ellipse). In that time interval, the TEC difference reaches amplitude of ~0.8 TECu centered at the EQ shock moment.

**Fig. 9b.** Non-interpolated vertical TEC difference TEC(untr) −TEC(m0se) for 5-8 April 2009 taken from all satellites crossing GPS stations in Central Italy with elevation angles exceeding 67 (blue) and 86 (red) degrees. Note that around the EQ shock moment there were no satellites with elevation angles exceeding 86 degrees. The TEC anomaly is 'caught' by satellites of less elevation angles (between 67 and 86 degrees). In a time interval of several hours, the TEC difference represents a *hump-shaped* distribution and reaches amplitude of ~0.8 TECu centered at the EQ shock moment.

**Fig. 10.** Cones of line-of-sights trajectories centered at two TEC stations: *m0se* and *Aqui*, EQ epicenter (marked with four-point star) and concentric fronts of seismogenic disturbances above the EQ epicenter (in grey) are illustrated. Disturbances in the ionosphere within the cones (marked with ellipses) become detached to each other only for heights less than 160 km. Above these heights the cones and associated disturbances become overlapping and hence cannot be easily separated.

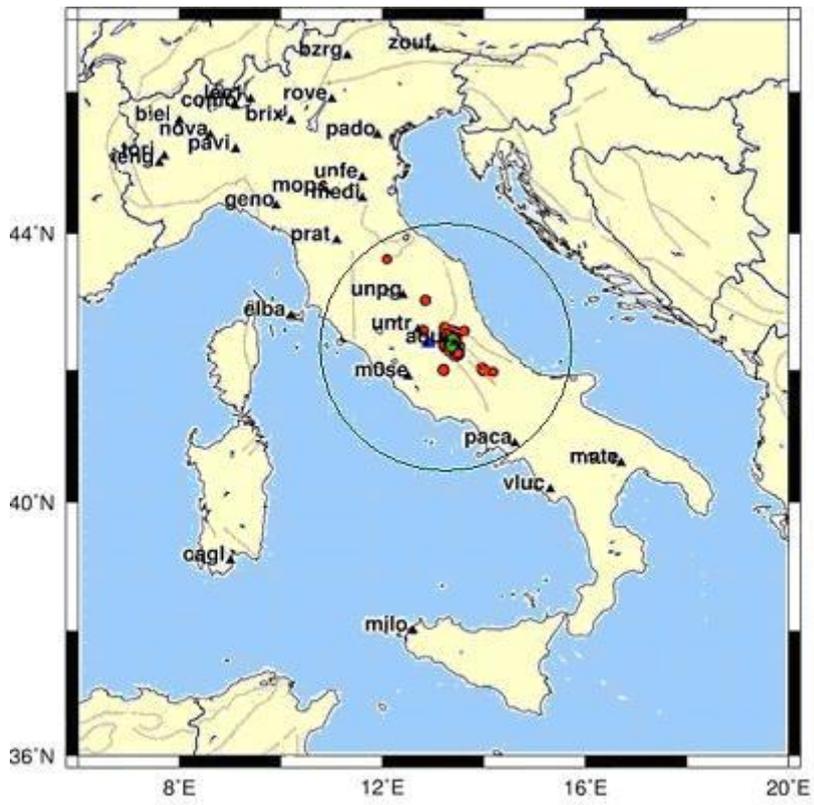

**Fig. 1.**

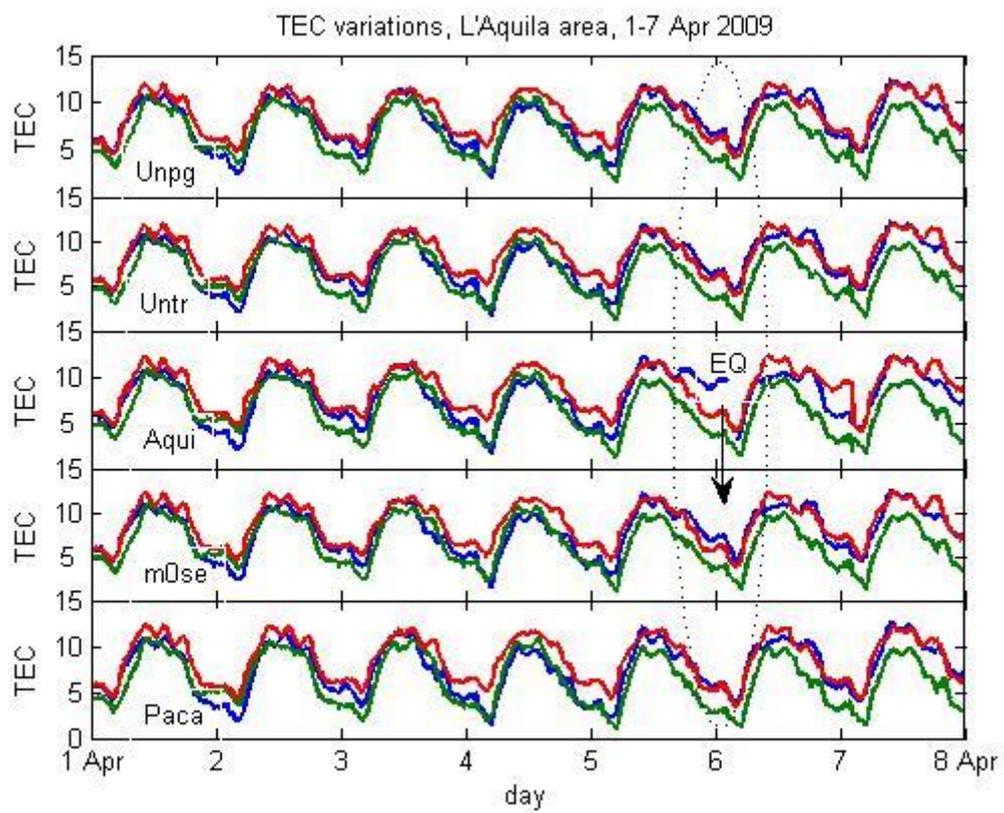

**Fig. 2.**

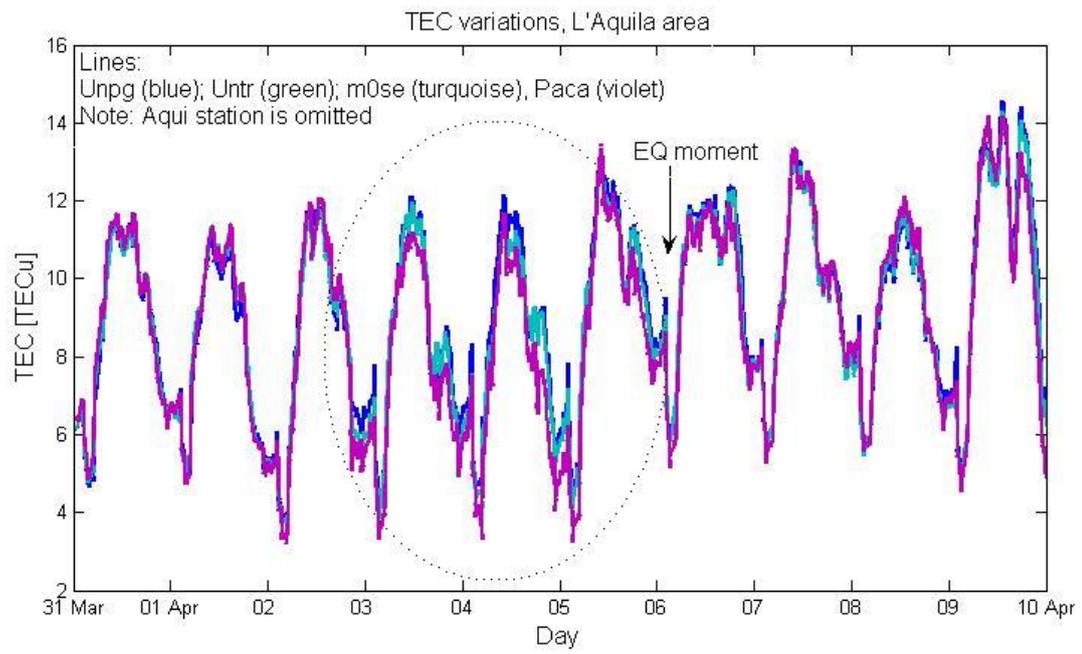

**Fig. 3.**

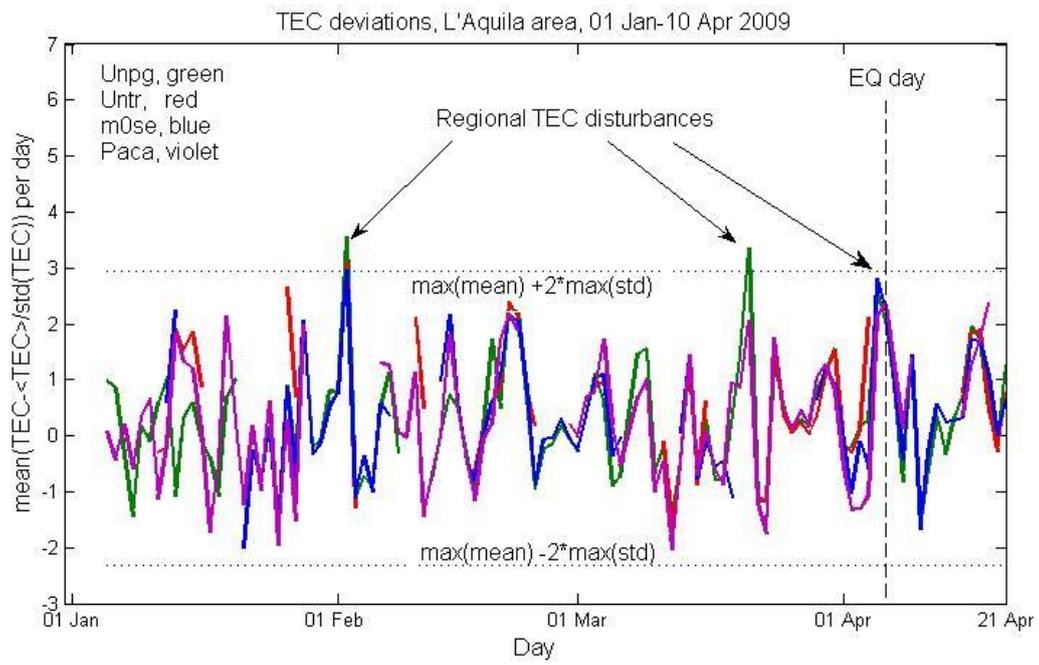

**Fig. 4.**

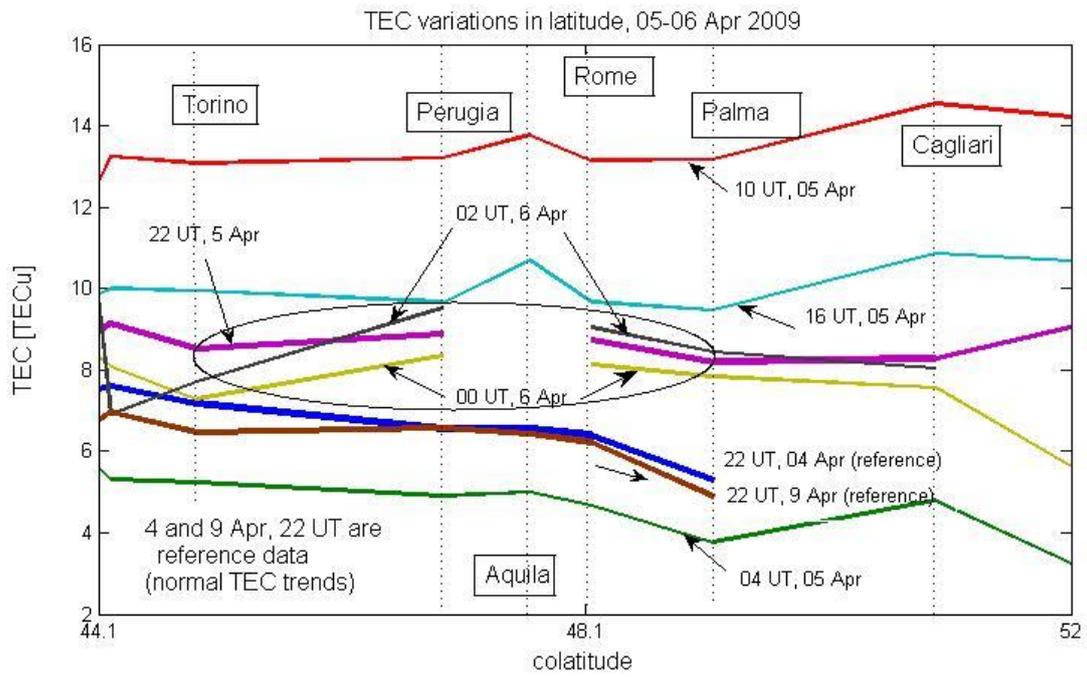

**Fig. 5.**

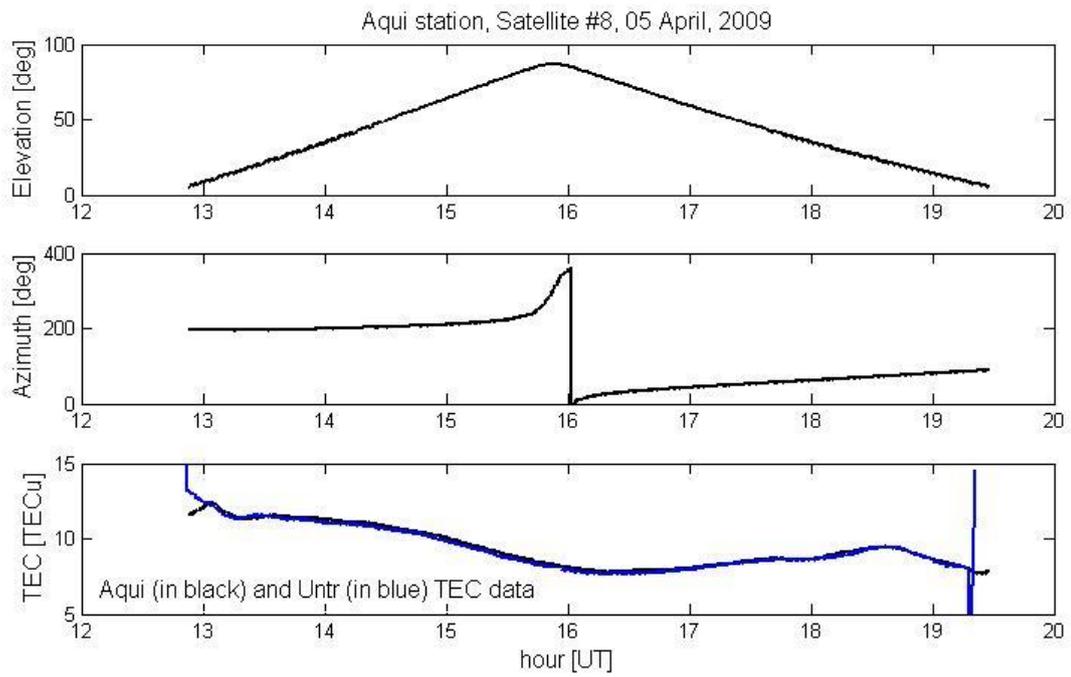

**Fig. 6.**

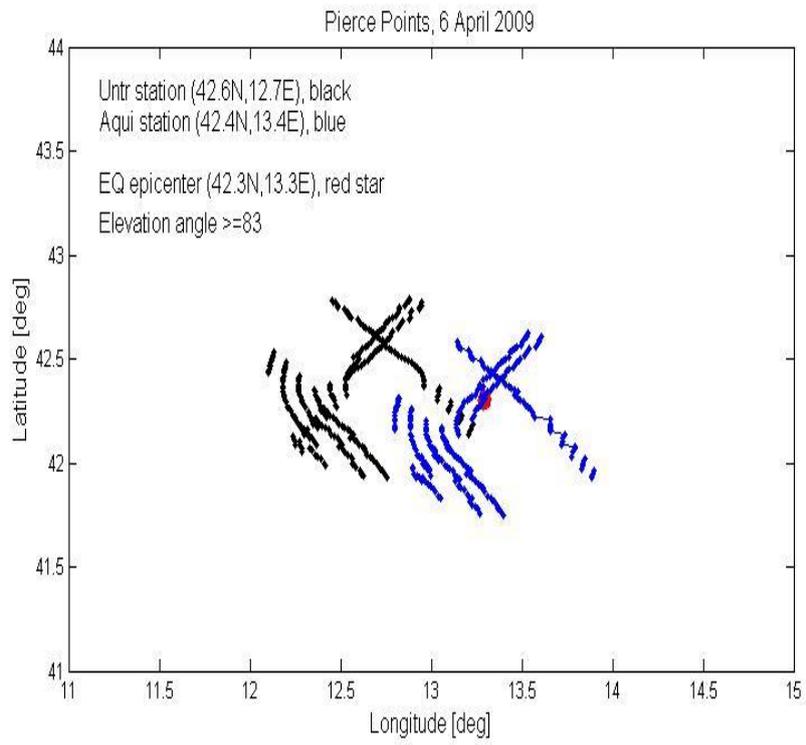

**Fig. 7**

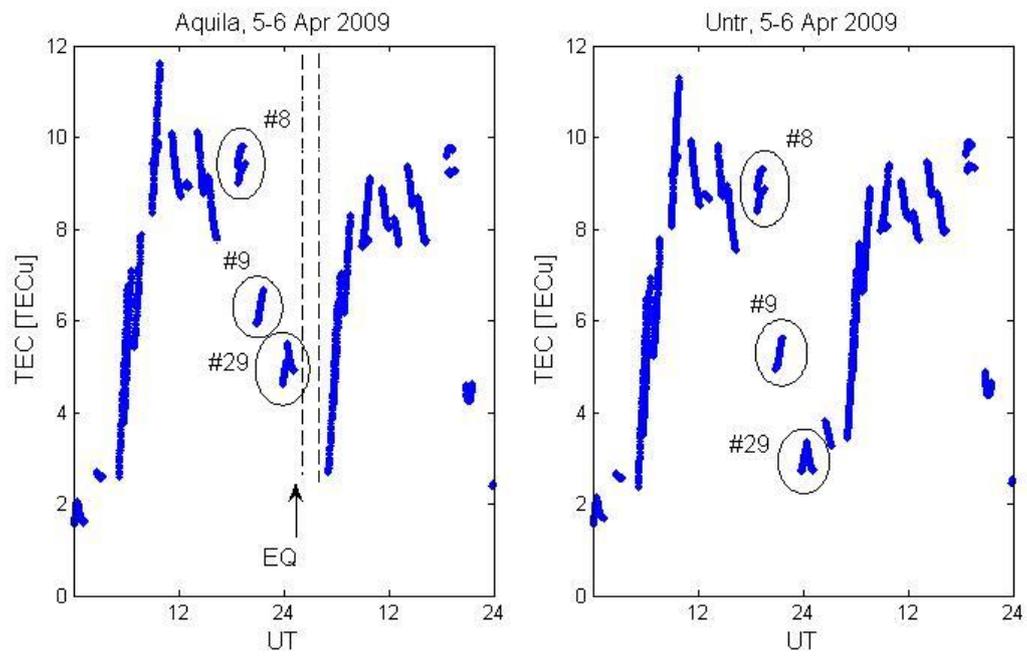

**Fig. 8a.**

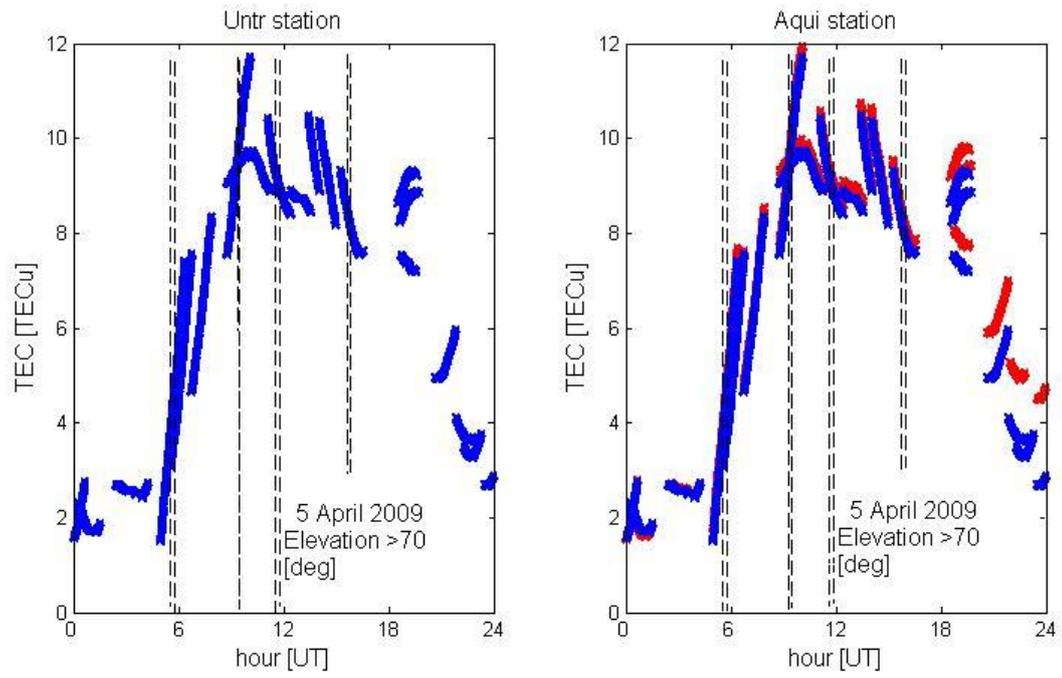

**Fig. 8b.**

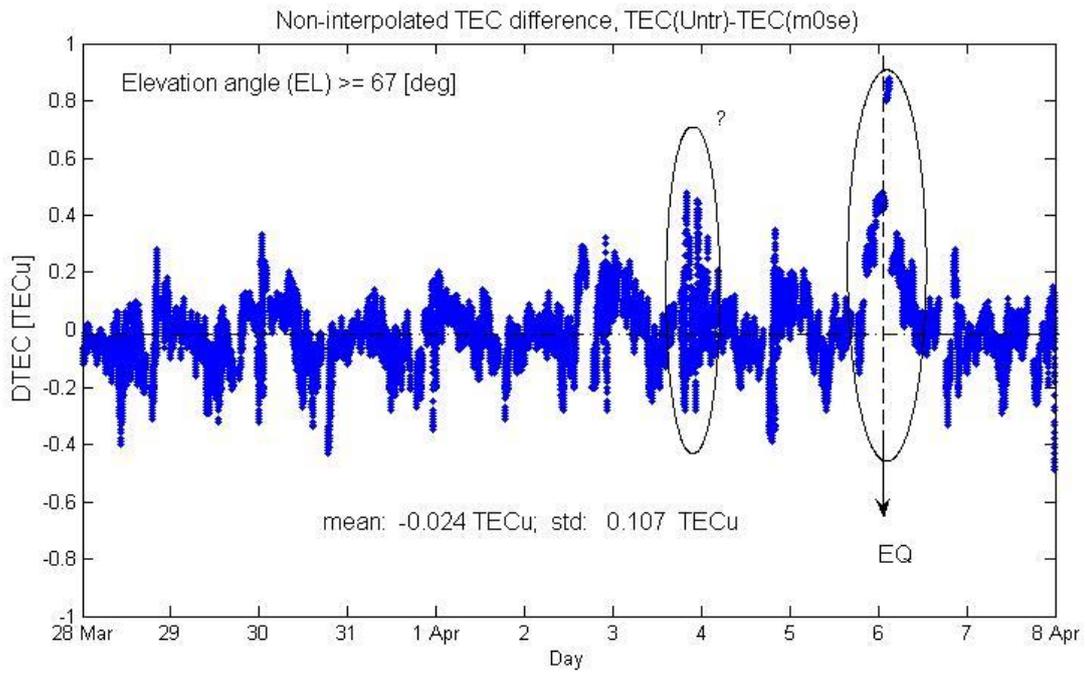

**Fig. 9a.**

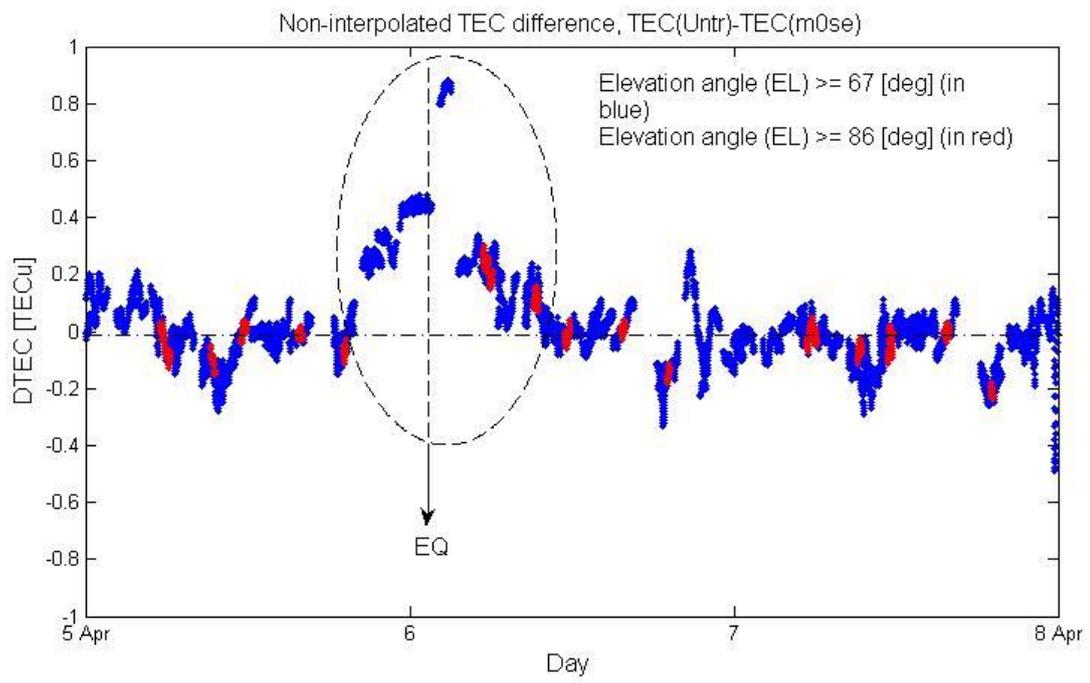

**Fig. 9b.**

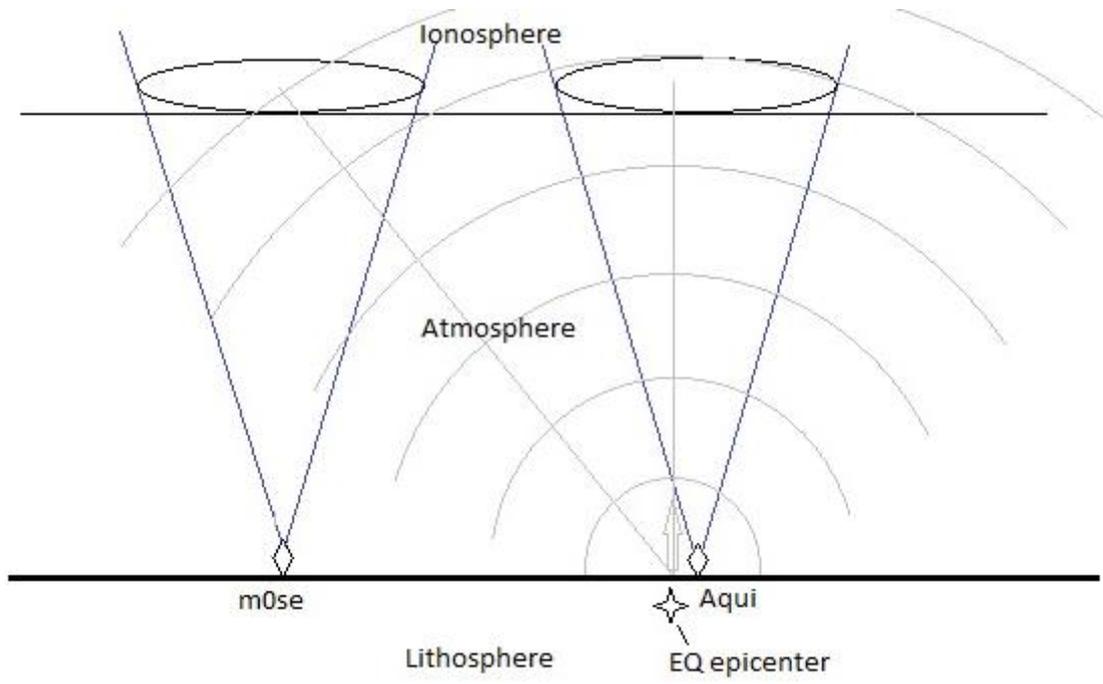

**Fig. 10.**